\documentclass[a4paper,fleqn,usenatbib,useAMS]{mnras}

\usepackage{graphicx}	
\usepackage{amsmath}	
\usepackage{amssymb}	
\usepackage{multicol}        
\usepackage{bm}		
\usepackage{pdflscape}	
\usepackage{multirow}
\usepackage{xcolor}
\usepackage{hyperref}
\usepackage{booktabs}
\usepackage{subfigure}
\usepackage{tikz}

\newcommand{\kms}{\,km\,s$^{-1}$} 
\def\specialname[#1]{\textbf{\textsc{#1}}}

\definecolor{lime}{HTML}{A6CE39}
\DeclareRobustCommand{\orcidicon}{%
	\begin{tikzpicture}
	\draw[lime, fill=lime] (0,0) 
	circle [radius=0.16] 
	node[white] {{\fontfamily{qag}\selectfont \tiny ID}};
	\draw[white, fill=white] (-0.0625,0.095) 
	circle [radius=0.007];
	\end{tikzpicture}
	\hspace{-2mm}
}
\foreach \x in {A, ..., Z}{%
	\expandafter\xdef\csname orcid\x\endcsname{\noexpand\href{https://orcid.org/\csname orcidauthor\x\endcsname}{\noexpand\orcidicon}}
}

\usepackage[T1]{fontenc}
\usepackage{ae,aecompl}

\usepackage{newtxtext,newtxmath}

\title[Dissect two-halo galactic conformity]{
    Dissect two-halo galactic conformity effect for central galaxies:
    The dependence of star formation activities on the large-scale environment
}
\author[Wang et al.]{
    Kai Wang\orcidK{},$^{1}$\thanks{e-mail: wkcosmology@gmail.com\href{}{}}
	Yingjie Peng$^{2, 1}$\thanks{e-mail: yjpeng@pku.edu.cn}
    and Yangyao Chen\orcidC{}$^{3, 4}$
	\\
    $^{1}$Kavli Institute for Astronomy and Astrophysics, Peking University, Beijing 100871, China\\
    $^{2}$Department of Astronomy, School of Physics, Peking University, Beijing 100871, China \\
    $^{3}$School of Astronomy and Space Science, University of Science and Technology of China, Hefei 230026, China\\
    $^{4}$Key Laboratory for Research in Galaxies and Cosmology, Department of Astronomy, University of Science and Technology of China, Hefei, Anhui 230026, China}

\date{Last updated 2020 May 22; in original form 2018 September 5}

\pubyear{2020}

\begin{document}
	\label{firstpage}
	\pagerange{\pageref{firstpage}--\pageref{lastpage}}
	\maketitle


\begin{abstract}
    We investigate the two-halo galactic conformity effect for
    central galaxies, which is the spatial correlation of the star formation
    activities for central galaxies to several Mpcs, by studying the dependence
    of the star formation activities of central galaxies on their large-scale
    structure in our local Universe using the SDSS data. Here we adopt a novel
    environment metric using only central galaxies quantified by the distance
    to the $n$-th nearest central galaxy. This metric measures the environment
    within an aperture from $\sim 1$ Mpc to $\gtrsim 10$ Mpc, with a median
    value of $\sim 4$ Mpc. We found that two kinds of conformity effects in our
    local Universe. The first one is that low-mass central galaxies are more
    quenched in high-density regions, and we found that this effect mainly
    comes from low-mass centrals that are close to a more massive halo. A
    similar trend is also found in the IllustrisTNG simulation, which can be
    entirely explained by backsplash galaxies. The second conformity effect is
    that massive central galaxies in low-density regions are more star-forming.
    This population of galaxies also possesses a higher fraction of spiral
    morphology and lower central stellar velocity dispersion, suggesting that
    their low quiescent fraction is due to less-frequent major merger events
    experienced in the low-density regions, and as a consequence, less-massive
    bulges and central black holes.
\end{abstract}

\begin{keywords}
	methods: statistical - galaxies: groups: general - dark matter - large-scale structure of Universe
\end{keywords}



\section{Introduction}%
\label{sec:introduction}

Observational data of galaxy surveys shows a strong bimodality in terms of
galaxy properties, where one population is blue, star-forming, and spiral-like,
while the other one is red, quiescent, and spheroidal-like
\citep[e.g.][]{stratevaColorSeparationGalaxy2001, hoggLuminosityDensityRed2002,
    ellisMillenniumGalaxyCatalogue2005, ballBivariateGalaxyLuminosity2006,
baldryGalaxyBimodalityStellar2006a, pengMASSENVIRONMENTDRIVERS2010}. Galaxies
must be star-forming at first to accumulate their stellar mass before evolving
into the quiescent population, where this transition is also known as the
quenching process. There are many quenching mechanisms proposed in the
literature, and they can be categorized into internal and external channels.
The former includes the feedback effects from the accretion of central massive
black holes \citep[e.g.][]{crotonManyLivesActive2006,
fabianObservationalEvidenceAGN2012}, the inefficient cooling for gas in massive
halos \citep{dekelGalaxyBimodalityDue2006}, morphological quenching
\citep{martigMorphologicalQuenchingStar2009, genzelSINSZCSINFSurvey2014,
gensiorHeartDarknessInfluence2020}, and angular momentum quenching
\citep{renziniAngularMomentumHistory2020, pengDiscGrowthQuenching2020}.
Meanwhile, the latter includes the strangulation effect
\citep[e.g.][]{larsonEvolutionDiskGalaxies1980,
pengStrangulationPrimaryMechanism2015}, the harassment effect
\citep[e.g.][]{mooreGalaxyHarassmentEvolution1996,
smithHowEffectiveHarassment2010}, and the ram-pressure stripping effect
\citep{gunnInfallMatterClusters1972a}. In addition,
\citet{pengMASSENVIRONMENTDRIVERS2010} showed that the dependence of the
quiescent fraction on stellar mass and local environment is empirically
separable, suggesting that these two mechanisms, i.e. mass quenching and
environment quenching, are affecting the star formation activities of galaxies
independently \citep[see also][]{baldryGalaxyBimodalityStellar2006a}.
Furthermore, \citet{pengMASSENVIRONMENTDRIVERS2012} found that the quiescent
fraction of central galaxies primarily depends on stellar mass, while both mass
quenching and environment quenching are affecting the evolution of satellite
galaxies \citep[see also][]{wangELUCIDIVGalaxy2018}.

In addition, \citet{weinmannPropertiesGalaxyGroups2006} found that the
star-forming activities of satellites also correlate with the star formation
states of their central galaxies with both stellar mass and halo mass fixed.
This effect is known as the one-halo conformity effect \citep[see
also][]{knobelQUENCHINGSTARFORMATION2015, treyerGroupQuenchingGalactic2018}.
The robustness of this effect is still under debate since the halo mass
calibration through the heuristic abundance matching method ignores their
dependence on the colors/star-forming activities of central galaxies
\citep{mandelbaumStrongBimodalityHost2016}. Later,
\citet{kauffmannReexaminationGalacticConformity2013} found that the correlation
between the star formation states of central galaxies and their neighboring
galaxies extends to several Mpc, much larger than the virial radius of host
halos. This effect is named the two-halo galactic conformity effect \citep[see
also][]{hearinHaloMassGalactic2015, hearinPhysicalOriginGalactic2016,
sinGeneralApproachQuenching2019, ayromlouPhysicalOriginGalactic2022a}, which is
the focus of this paper. It is noteworthy that similar conformity effects are
also found for other properties of galaxies
\citep[e.g.][]{wangViewGalaxyConformity2015a,
calderonSmallLargescaleGalactic2018, otterGalacticConformityBoth2020} and for
high-z galaxies \citep[e.g.][]{hartleyGalacticConformityCentral2015,
kawinwanichakijSatelliteQuenchingGalactic2016, bertiPRIMUSOneTwohalo2017}.

There are many mechanisms proposed in the literature to explain the two-halo
conformity effect. \citet{kauffmannPhysicalOriginLargescale2015} proposed that
gas is heated over large scales by the active galactic nuclei (AGN) feedback
effect, since they found an excess number of massive galaxies around quiescent
central galaxies, and these massive galaxies have higher probability to host
radio-loud AGNs. \citet{hearinHaloMassGalactic2015} claimed that the conformity
signal is a smoking gun of halo assembly bias, i.e. early-formed halos are more
clustered, since they can reproduce the conformity signal in simulations
assuming that central galaxies in early-formed halos are more quenched
\citep[see also][]{wangRelatingGalaxiesDifferent2023}. Based on the same
philosophy, \citet{hearinPhysicalOriginGalactic2016} found that the conformity
signal can also be explained by correlating the star-forming activities of
central galaxies with the accretion rate of their host halos, since the latter
exhibits strong spatial correlation over several Mpcs. Alternatively,
\citet{zuMappingStellarContent2018} showed that the conformity signal can be
reproduced in a model that galaxy quenching is only determined by their stellar
mass and host halo mass, and no super-halo-scale physical processes nor
assembly bias is required. Recently,
\citet{ayromlouPhysicalOriginGalactic2022a} found that the conformity effect
may originate from the super-halo-scale ram-pressure stripping effect. The
evidence is that the conformity signal emerges from the semi-analytical model
of \texttt{L-GALAXIES} \citep{henriquesGalaxyFormationPlanck2015,
henriquesLGALAXIES2020Spatially2020} once a recipe that can strip gas from
galaxies out of halo boundaries is introduced
\citep{ayromlouGalaxyFormationLGALAXIES2021}.

The two-halo galactic conformity effect also manifests itself in the
semi-analytical and hydrodynamical models of galaxy formation
\citep{brayModellingGalacticConformity2016,
lacernaGalacticConformityMeasured2018}.
\citet{sunLargescaleEffectEnvironment2018} showed that the two-halo galactic
conformity effect is due to that the star formation activities of central
galaxies have a dependence on their environment, and they found a similar
signal in the semi-analytical model of \citet{guoDwarfSpheroidalsCD2011}.
Besides, \citet{lacernaEnvironmentalInfluenceGroups2022} found similar two-halo
conformity signals in the semi-analytical model of \texttt{SAG} and the
IllustrisTNG hydrodynamical simulation, and this signal can be eliminated once
central galaxies in the vicinity of massive clusters are removed
\citep{sinEvidenceLargeScaleGalactic2017,
ayromlouGalaxyFormationLGALAXIES2021}.

The original definition of two-halo galactic conformity in
\citet{kauffmannReexaminationGalacticConformity2013} is the correlation between
the star formation in central galaxies (primary galaxies), which are identified
using the isolation criterion, and their neighbors (secondary galaxies) up to
several Mpcs. This definition has several flaws. First, the isolation criterion
can introduce fake signals by misidentifying satellite galaxies as centrals
\citep{sinEvidenceLargeScaleGalactic2017, tinkerHaloHistoriesGalaxy2018}.
Second, the secondary galaxies include both central and satellite galaxies, so
that the one-halo and two-halo galactic conformity effects are mixed, which
makes the result hard to interpret \citep[][see also
\S\,\ref{sub:two_halo_conformity_effect_on_satellite_galaxies}]
{paranjapeCorrelatingGalaxyColour2015, tinkerHaloHistoriesGalaxy2017,
tinkerHaloHistoriesGalaxy2018a}. Finally, the correlation between primary and
secondary galaxies can be explained by either a causal relation between them,
or a scenario that both kinds of galaxies are affected by a mediate property,
or both. In order to alleviate the impact of these flaws and make the results
interpretable, here we study the star formation of central galaxies, identified
by the halo-based group finding algorithm of \citet{yangGalaxyGroupsSDSS2007}
\citep[see also][]{yangHalobasedGalaxyGroup2005, luGALAXYGROUPS2MASS2016,
    limGalaxyGroupsLowredshift2017, wangIdentifyingGalaxyGroups2020,
yangExtendedHalobasedGroup2021, liGroupsProtoclusterCandidates2022}, and its
dependence on the large-scale environment traced by central galaxies only. In
the remaining content, we will refer to the dependence of the star formation
activities of central galaxies on the large-scale environment as the two-halo
conformity effect, unless specified otherwise.

This paper is organized as follows. \S~\ref{sec:data} introduces the galaxy
sample from observation and simulation, as well as the calculation of our
large-scale environment metric. The main results are presented in
\S~\ref{sec:results}, and the discussion is in \S~\ref{sec:discussion}. And we
will summarize in \S~\ref{sec:conclusion}. Throughout this paper, we are
assuming a concordance $\Lambda$CDM cosmology with $H_0=100h~\rm km/s/Mpc$,
$h=0.7$, $\Omega_{\Lambda}=0.75$, and $\Omega_{m}=0.25$.

\section{Data}%
\label{sec:data}

\begin{figure*}
    \centering
    \includegraphics[width=1\linewidth]{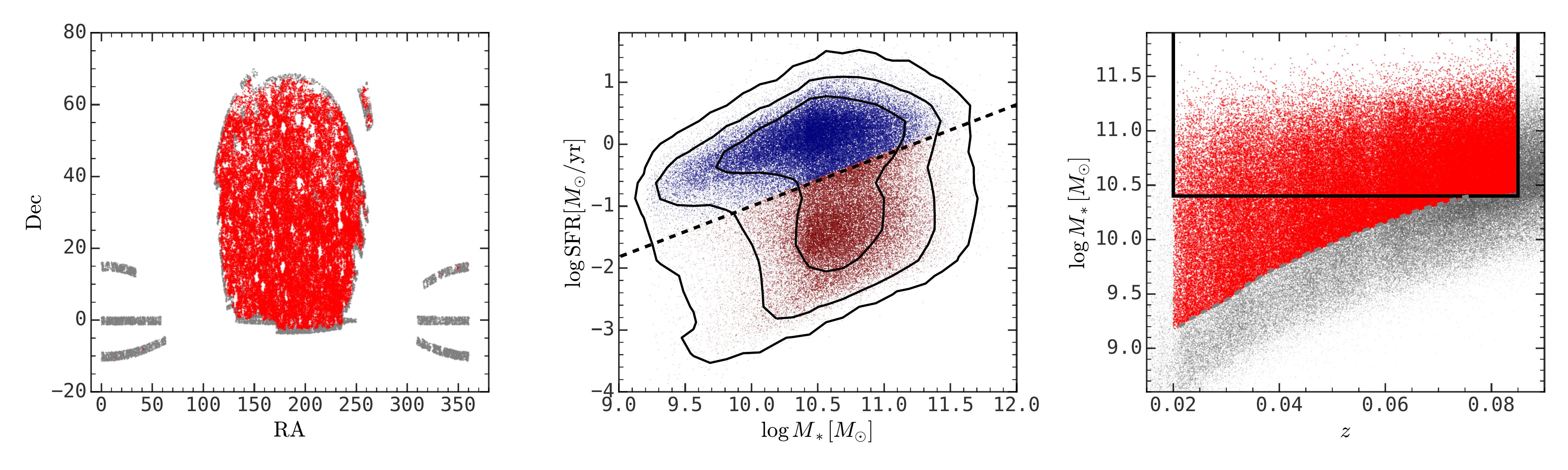}
    \caption{
        {\bf Left panel:} The projected distribution of the SDSS main galaxy
        sample, where red points are the selected mass-limited galaxies with
        $f_{\rm 10Mpc} > 0.9$ and gray points are the remaining ones. {\bf
        Middle panel:} The distribution of galaxies on the stellar mass and SFR
        plane with black solid lines enclose 68\%, 95\%, and 99.7\% of the
        whole sample. The dashed line is the calibrated separation line for the
        star-forming and quiescent galaxies. The red points are quiescent
        galaxies and the blue ones are star-forming galaxies. {\bf Right
        panel:} The distribution of galaxies on the redshift and stellar mass
        plane, where red points are the selected mass-limited galaxies with
        $f_{\rm 10Mpc} > 0.9$ and gray points are the remaining ones. The black
        solid lines enclose a volume-limited sample with $M_* >
        10^{10.4}M_{\odot}$ and $0.02 < z < 0.09$. The central galaxies in the
        volume-limited sample, i.e. those enclosed by the black solid box, are
        environment definition tracers, while galaxies in the mass-limited
        sample, i.e. those above the black dashed line, are used for analysis,
        such as calculating the quiescent fraction.
    }%
    \label{fig:figure/sdss_vlim_sample}
\end{figure*}

\begin{figure*}
    \centering
    \includegraphics[width=1\linewidth]{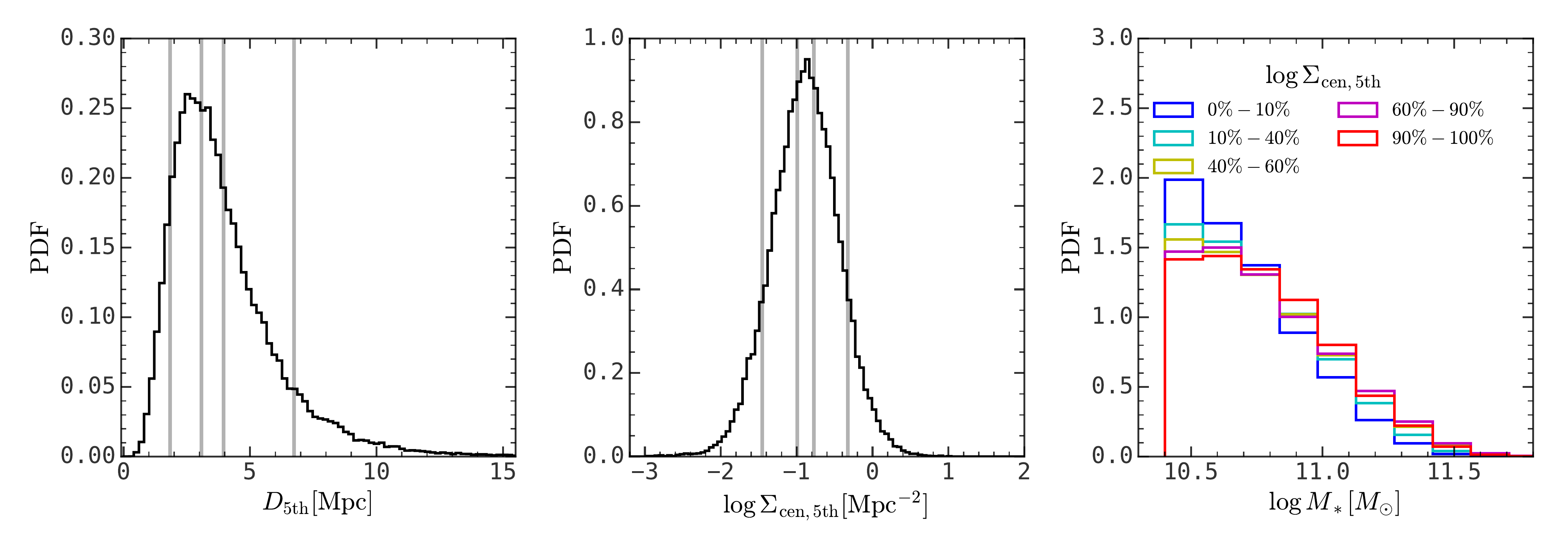}
    \caption{
        {\bf Left panel:} The distribution of the projected distance to the 5th
        nearest central galaxy within $\pm 1000\rm km/s$, i.e. $D_{\rm 5th}$,
        for the volume-limited sample, i.e. those in enclosed the black solid
        box on the left panel of Fig.~\ref{fig:figure/sdss_vlim_sample}. {\bf
        Middle panel:} The distribution of the surface number density, i.e.
        $\Sigma_{\rm cen, 5th}$, calculated with equation~(\ref{eq:sigma_cen})
        for the volume-limited sample. {\bf Right panel:} The distribution of
        stellar mass in different density quantiles for the volume-limited
        sample. The vertical lines in the first two panels correspond to the
        10\%, 40\%, 60\%, and 90\% quantiles of $\Sigma_{\rm cen, 5th}$ for all
        the mass-complete central galaxies. The vertical gray lines on the
        first two panels are the 10\%, 40\%, 60\%, and 90\% percentiles.
    }%
    \label{fig:figure/sdss_density_distribution_cen}
\end{figure*}

\subsection{Galaxy catalog}%
\label{sub:galaxy_catalog}

The sample used for this study is drawn from the Sloan Digital Sky Survey
(SDSS) main galaxy sample (MGS), which is a magnitude-limited spectroscopic
survey with $r < 17$ covering $\sim 8000~\rm deg^2$ carried on the 2.5 meter
wide-angle telescope at Apache Point Observatory
\citep{yorkSloanDigitalSky2000, blantonNewYorkUniversity2005,
abazajianSeventhDataRelease2009}. The SDSS MGS catalogue comprises five bands
photometry ($u$, $g$, $r$, $i$, $z$) and accurate redshift measurements from
its spectroscopic observation. The central stellar velocity dispersion is
measured within the $3^{\prime\prime}$ diameter fiber. Derived quantities, like
stellar mass and star formation rate (SFR), are obtained from multi-band
photometric measurements and spectra indices
\citep{kauffmannStellarMassesStar2003, salimGALEXSDSSWISE2016}. In the present
work, we adopt the stellar mass and SFR from the GALEX-SDSS-WISE Legacy Catalog
(GSWLC) \citep{salimGALEXSDSSWISE2016, salimDustAttenuationCurves2018}, which
are estimated by fitting the UV-optical-IR bands photometry using the CIGALE
code \citep{nollAnalysisGalaxySpectral2009, boquienCIGALEPythonCode2019} with
the stellar library of \citet{bruzualStellarPopulationSynthesis2003} and the
initial mass function of \citet{chabrierGalacticStellarSubstellar2003}. The
specific star formation rate (SSFR) is defined as $\rm SFR/M_*$. The middle
panel of Fig.~\ref{fig:figure/sdss_vlim_sample} shows the distribution of
galaxies on the $M_*-\rm SFR$ plane where one can clearly see the bimodality.
We further separate galaxies into star-forming and quiescent populations,
according to the method in \citet{wooDependenceGalaxyQuenching2013}. We first
fit a linear function to the star-forming main sequence by iteratively applying
the least-square fitting and excluding data points 1 dex below the main
sequence until convergence. Finally, we define galaxies with SFR lower than the
main sequence by more than 1 dex as quiescent, and the remaining ones as
star-forming. The separation line is
\begin{equation}
    \log \left(\frac{\rm SFR}{M_{\odot}/{\rm yr}}\right) = 0.82\times \log
    \left(\frac{M_*}{M_{\odot}}\right) - 9.17,
\end{equation}
and shown as the black dashed line in the middle panel of
Fig.~\ref{fig:figure/sdss_vlim_sample}.

Directly calculating the quiescent fraction in stellar mass bins for a
magnitude-limited sample, like a $r$-band limited sample, will produce biased
results, since star-forming galaxies are brighter in the blue band and hence
they are preferentially selected, compared with their quiescent counterparts
with similar stellar mass. In previous studies, a $K$-correction procedure is
applied to all galaxies to calculate the $V_{\rm max}$ factors
\citep{blantonKCorrectionsFilterTransformations2007}, which are used for
weighting these galaxies \citep[e.g.][]{pengMASSENVIRONMENTDRIVERS2010}. A more
conservative method is to construct a $M_*$-limited sample, which contains a
complete sample of galaxies above some stellar mass threshold, which is a
function of redshift \citep[see][for more
details]{pozzettiZCOSMOS10kbrightSpectroscopic2010}, after which we can
calculate a $V_{\rm max}$ factor for each galaxy in the mass-limited sample. We
adopted the latter in this paper. To begin with, we divide galaxies into small
redshift bins, i.e. $\Delta z = 0.02$. Then, in each redshift bin, we select
20\% galaxies with faintest $r$-band magnitude and calculate the rescaled
stellar mass, i.e.
\begin{equation}
    \log \left(\frac{M_{\rm scale}}{M_{\odot}}\right) = \log
    \left(\frac{M_*}{M_{\odot}}\right) + 0.4 \times (r - r_{\rm lim})
\end{equation}
where $M_*$ and $r$ are the stellar mass and $r$-band apparent magnitude, and
$r_{\rm lim}$ is the limit $r$-band magnitude, which ranges from 17.57 to 17.72
for SDSS galaxies. Finally, the stellar mass limit is the envelope assembled by
the 95\% percentile of $M_{\rm scale}$ in each redshift bin. The envelope is
shown as the gray dashed line on the right panel of
Fig.~\ref{fig:figure/sdss_vlim_sample} and the mass-limited sample is shown in
red points. We also weigh each galaxy with $1/V_{\rm max}$ where $V_{\rm max}$
is the volume between $z_{\rm min}$ and $z_{\rm max}$ with $z_{\rm min} = 0.02$
and $z_{\rm max}$ inferred from the mass-limited envelop calculated above
according to the stellar mass of each galaxy \citep[see
also][]{2023arXiv230407189W}. In addition, we constructed a volume-limited
sample by selecting galaxies with
\begin{equation}
    M_* > 10^{10.4}M_{\odot}~~~\&~~~0.02 < z < 0.09. \label{eq:mass_complete}
\end{equation}
We emphasize that the central galaxies in the mass-limited sample are the
objects for investigation in this work, and the central galaxies in the
volume-limited sample only serve as environment tracers, which will be
introduced in \S\,\ref{sub:large_scale_environment}.

We use the morphology classification from the Galaxy Zoo
project\footnote{https://data.galaxyzoo.org/} \citep{lintottGalaxyZooData2011}.
In Galaxy Zoo, galaxies are classified as \texttt{spiral}, \texttt{elliptical},
and \texttt{uncertain} according to their image, and a debiasing process is
applied. Here we define spiral galaxies as ones with the flag of
\texttt{SPIRAL} set to one.

We use the group catalog of \citet{yangGalaxyGroupsSDSS2007} to select the
central galaxy as the most massive one in each galaxy group. The halo mass is
calibrated by abundance matching the total stellar mass with the theoretical
halo mass function. We note that some studies require the central galaxy to be
the most massive and the brightest one simultaneously. This more strict
criterion only eliminates $\sim 2\%$ central galaxies from our sample
\citep{pengMASSENVIRONMENTDRIVERS2012}, and has negligible impact on the
results presented in this paper.

\subsection{Simulation Data}%
\label{sub:simulation_data}

The IllustrisTNG project \citep{pillepichSimulatingGalaxyFormation2018,
    springelFirstResultsIllustrisTNG2018, naimanFirstResultsIllustrisTNG2018,
    marinacciFirstResultsIllustrisTNG2018,
    pillepichFirstResultsIllustrisTNG2018, nelsonFirstResultsIllustrisTNG2018,
nelsonIllustrisTNGSimulationsPublic2019} comprises a set of
gravo-magnetohydrodynamical cosmological simulations that run with the
moving mesh code \texttt{AREPO} \citep{springelPurSiMuove2010}. It
simulates the formation and evolution of galaxies from $z=127$ to $z=0$
based on a cosmology constrained in
\citet{planckcollaborationPlanck2015Results2016}, where $\Omega_{\Lambda,
0} = 0.6911$, $\Omega_{b, 0}=0.3089$, $\sigma_8=0.8159$, $n_s=0.9667$, and
$h=0.6774$. In this paper, we are using the version with a box side length
of $\sim 300$Mpc for better statistics. This simulation contains $2500^3$
dark matter particles with each weights $\sim5.9\times 10^7\rm M_{\odot}$,
and $\sim 2500^3$ gas cells with each weights $\sim 1.1\times 10^7\rm
M_{\odot}$.

The dark matter halos are identified with the friends-of-friends (FoF)
algorithm \citep{davisEvolutionLargescaleStructure1985} using dark matter
particles. Substructures are identified with the \texttt{SUBFIND} algorithm
using all types of particles \citep{springelPopulatingClusterGalaxies2001},
then the baryonic components are defined as galaxies and the dark matter
components are defined as subhalos. In each FoF halo, the central subhalo is
defined as the one that is located at the minimum of the gravitational
potential, and the galaxy inside this subhalo is defined as the central galaxy,
while the remaining subhalos and galaxies are satellites. Subhalo merger trees
are constructed using the \texttt{SUBLINK} algorithm
\citep{rodriguez-gomezMergerRateGalaxies2015}, and we identify the main
progenitor for each subhalo/galaxy as the one with {\it the most massive
progenitor history} \citep{deluciaHierarchicalFormationBrightest2007}.

Here we define the halo mass for each FoF halo as the total mass within a
radius within which the average density equals 200 times the critical density,
while the radius is used as the halo radius and denoted as $R_{200}$. The
stellar mass is calculated by summing all the stellar particle mass within
$2R_*$, where $R_*$ is the stellar half-mass radius for all the stellar
particles attributed to the subhalo. The SFR is also calculated from all the
star-forming particles within $2R_*$. For galaxies in the IllustrisTNG
simulation, those with $\rm SSFR < 10^{-11}\rm yr^{-1}$ are defined as
quiescent, while the remaining ones are star-forming. It is noteworthy that we
are using a different criterion to separate star-forming/quiescent galaxies in
observation and simulation. Since we do not intend to perform a quantitative
comparison between simulation and observation
\citep[e.g.][]{donnariStarFormationActivity2019,
donnariQuenchedFractionsIllustrisTNG2021}, it is fine as long as this
separation criterion does not bias the environmental dependence of the
quiescent fraction.

For all central galaxies at $z=0$, we also identify a sub-sample of backsplash
galaxies with the following properties
\citep{wangDistributionEjectedSubhaloes2009, wetzelGalaxyEvolutionGroups2014}:
\begin{enumerate}
    \item They are central galaxies at $z=0$.
    \item Their main progenitors were once satellites of other halos and the
        satellite state lasts at least two successive snapshots. Here the
        second requirement is to avoid the {\it central-satellite switching
        problem} \citep[see Figure 1 in
        ][]{pooleConvergencePropertiesHalo2017}.
\end{enumerate}

\subsection{Large-scale environment}%
\label{sub:large_scale_environment}

The formation and evolution of galaxies and their associated dark matter halos
are subject to environmental effects \citep{moGalaxyFormationEvolution2010}.
There are many different metrics to quantify galaxy environments on different
spatial scales \citep[e.g.][]{muldrewMeasuresGalaxyEnvironment2012}. Here we
define a novel environment metric that only uses central galaxies as tracers,
which is defined as
\begin{equation}
    \Sigma_{\rm cen, 5th} = \frac{5}{\pi D_{\rm 5th}^2} \label{eq:sigma_cen}
\end{equation}
where $D_{\rm 5th}$ is the projected distance to the fifth nearest central
galaxy within $\pm 1000\rm km/s$ in the volume-limited galaxy sample. And each
central galaxy in the mass-limited sample has a value of $\Sigma_{\rm cen,
5th}$ calculated. The choice of the fifth neighbor is motivated by two reasons.
On one side, it cannot be too small which will suffer from the shot noise. On
the other side, it cannot be too large where all the information will be
smoothed out. Finally, we choose to use the fifth nearest neighbor.

The search of the nearest neighbors can be seriously biased near the edge of
survey regions and bright star masks. For the validity of our environment
metric, we exclude central galaxies whose 5th nearest neighbor searching is
biased. We randomly drop one million points in a rectangular region that
enclose the survey footprint on the sphere. Then, we calculate a completeness
factor for each central galaxy, which is defined as
\begin{equation}
    f_{R} = \frac{N(<\theta)}{\pi \theta^2\bar \Sigma},~~~\theta = R/D_A
\end{equation}
where $N(<\theta)$ is the number of random points within the survey mask and
within an angular distance of $\theta$ from the central galaxy in question
\citep{blantonNewYorkUniversity2005}, and $\bar \Sigma$ is the mean surface
density of random points, and $D_A$ is the angular distance of the central
galaxy in question. In the present work, we restrict our study to central
galaxies with $f_{\rm 10Mpc} > 0.9$, where 10Mpc is larger than the $D_{\rm
5th}$ for most of the central galaxies in this study, and we present the
projected distribution of these selected galaxies with red points on the left
panel of Fig.~\ref{fig:figure/sdss_vlim_sample}. We note that the spatial
incompleteness can bias the environment estimation by underestimating the local
density for galaxies near masked regions, so we make a conservative cut of 90\%
completeness to mitigate this effect. It is also noteworthy that this effect
can only undermine the environmental dependence of galaxy properties rather
than inducing it as long as the spatial mask does not correlate with the galaxy
properties in question.

In Fig.~\ref{fig:figure/sdss_density_distribution_cen}, we present the
distribution of $D_{\rm 5th}$ and $\Sigma_{\rm cen, 5th}$ for all central
galaxies in the volume-limited sample on the first two panels, with vertical
lines indicate the 10\%, 40\%, 60\%, and 90\% percentiles. One can see that
$D_{\rm 5th}$ spans a large range from $\lesssim 1\rm Mpc$ to $\gtrsim 10\rm
Mpc$. The middle panel shows the distribution of $\Sigma_{\rm cen, 5th}$. On
the right panel, we plot the stellar mass distributions for central galaxies in
the $M_*$-limited sample with different central-traced densities. One can clearly
see that the high-density regions are preferentially occupied by massive
galaxies.

\section{Results}%
\label{sec:results}

\begin{figure*}
    \centering
    \includegraphics[width=0.8\linewidth]{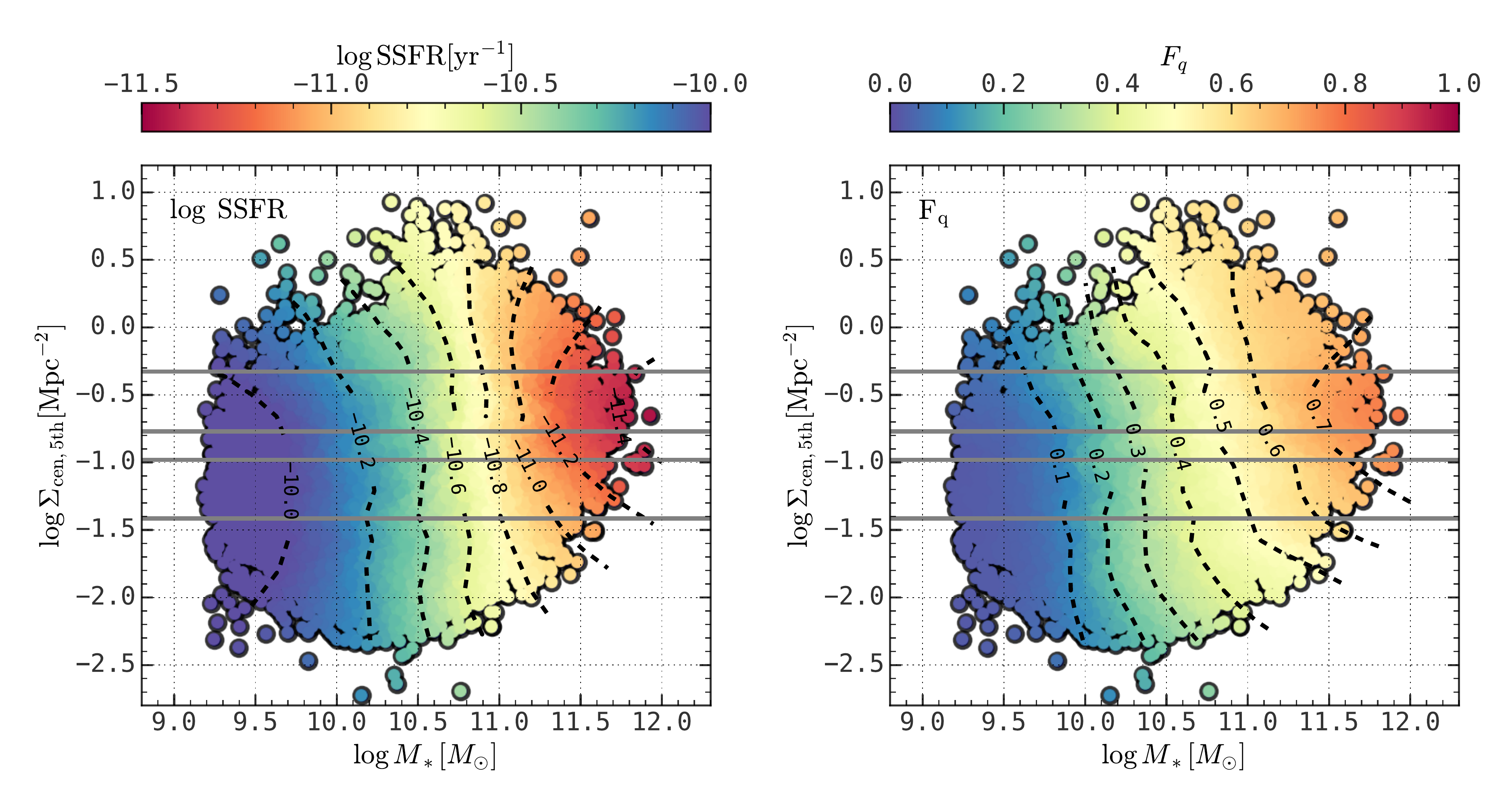}
    \caption{
        The median SSFR ({\bf left panel}) and quiescent fraction ({\bf right
        panel}) on the $\Sigma_{\rm cen, 5th}$ - stellar mass plane for SDSS
        galaxies, smoothed with the LOESS method. The black dashed lines are
        the contour lines and the horizontal gray solid lines are the 10\%,
        40\%, 60\%, and 90\% percentiles of $\Sigma_{\rm cen, 5th}$. For better
        visual illustration, we restrict the SSFR of galaxies to be within
        $(10^{-11.5},~10^{-10}){\rm yr}^{-1}$ on the right panel.
    }%
    \label{fig:figure/ssfr_fq_on_sm_sigma_cen_plane}
\end{figure*}

\begin{figure*}
    \centering
    \includegraphics[width=0.8\linewidth]{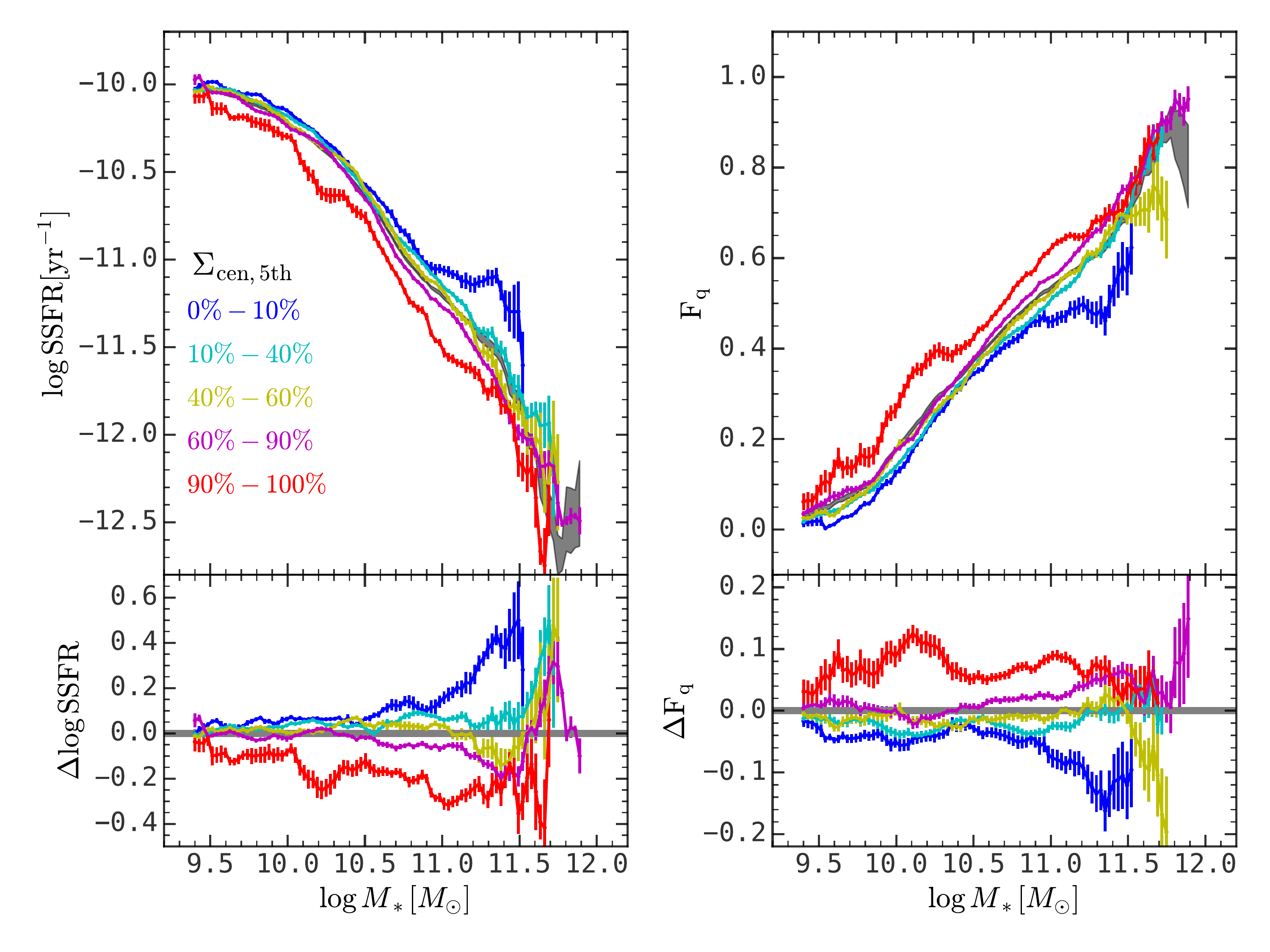}
    \caption{
        {\bf Top panels:} The median SSFR ({\bf left panel}) and the quiescent
        fraction ({\bf right panel}) as a function of stellar mass in different
        $\Sigma_{\rm cen, 5th}$ bins for SDSS galaxies. The gray regions are
        for all of the central galaxies. Error bars are calculated from
        bootstrap samples. {\bf Bottom panels:} The residual with respect to
        the result for all of the central galaxies, i.e. the gray regions on
        top panels.
    }%
    \label{fig:figure/gal_prop_cen_on_vmax_sigma_cen_1}
\end{figure*}

Fig.~\ref{fig:figure/ssfr_fq_on_sm_sigma_cen_plane} shows the median SSFR and
the quiescent fraction of central galaxies on the $M_*-\Sigma_{\rm cen, 5th}$
plane. The most obvious feature is that the gradient aligns with stellar mass,
consistent with previous works \citep{pengMASSENVIRONMENTDRIVERS2012}. For
low-mass central galaxies ($M_* \lesssim 10^{10.5}M_{\odot}$), one can see a
weak dependence on $\Sigma_{\rm cen, 5th}$ from the inclined contour lines,
especially above the 90\% quantile. This weak trend indicates that low-mass
central galaxies have lower SSFR and are more quenched in these dense regions.
Finally, the contour lines are also weakly inclined for massive galaxies ($M_*
\gtrsim 10^{11}M_{\odot}$) with low $\Sigma_{\rm cen, 5th}$, indicating that
these galaxies are more star-forming and less quenched than those with high
$\Sigma_{\rm cen, 5th}$. We note that these two trends are the manifestations
of the two-halo galactic conformity effect for central galaxies.

To see these signals more clearly,
Fig.~\ref{fig:figure/gal_prop_cen_on_vmax_sigma_cen_1} shows the median SSFR
and the quiescent fraction as a function of stellar mass in five $\Sigma_{\rm
cen, 5th}$ bins. From the right panel of
Fig.~\ref{fig:figure/sdss_density_distribution_cen}, one can see that dense
regions prefer to host massive galaxies, which is consistent with previous
results that massive galaxies are more clustered than low-mass ones
\citep{moAnalyticModelSpatial1996, liDependenceClusteringGalaxy2006,
liStellarMassStellar2013, wangClusteringPropertiesHalo2016}. Consequently,
galaxies in higher-density regions are on average more massive than those in
relatively low-density regions in the same finite stellar mass bin, and this
will induce an artificial dependence on $\Sigma_{\rm cen, 5th}$. Thus,
following the method of \citet{hartleyGalacticConformityCentral2015}, we
construct histograms with narrow stellar mass bins, i.e. $\Delta \log M_*=0.1$,
and calculate a weighting factor for each central galaxy, which is
\begin{equation}
    w_i^j = {N_i}/{N_i^j}
\end{equation}
where $w_i^j$ is the weight for a central galaxy in the $i$-th $M_*$ bin and
the $j$-th $\Sigma_{\rm cen, 5th}$ bin, $N_i$ is the number of central galaxies
in the $i$-th $M_*$ bin, and $N_i^j$ is the number of central galaxies in the
$i$-th $M_*$ bin and $j$-th $\Sigma_{\rm cen, 5th}$ bin simultaneously
\citep[see also][]{kawinwanichakijSatelliteQuenchingGalactic2016}.

In Fig.~\ref{fig:figure/gal_prop_cen_on_vmax_sigma_cen_1}, the gray shaded
regions show the results for central galaxies regardless of their $\Sigma_{\rm
cen, 5th}$, and solid lines with different colors show the results for central
galaxies in five $\Sigma_{\rm cen, 5th}$ bins with error bar calculated using
the bootstrap method. Meanwhile, the lower panels show the deviation from the
results for all central galaxies. As one can see, central galaxies in the
densest regions are $\sim 10$ percent more quenched than those in the lowest
$\Sigma_{\rm cen, 5th}$-bin, and the SSFR is also lower by $\sim 0.2$ dex.
Moreover, we also find that the galaxies with $M_* > 10^{11} M_{\odot}$ in the
lowest-density regions are less quenched than their counterparts in more dense
regions by $\sim 15$ percent, and the median SSFR is also lower by $\sim 0.4$
dex. Although the star formation of central galaxies only weakly depends on
$\Sigma_{\rm cen, 5th}$, this dependence is statistically significant, which
can be seen from the small error bars estimated with the bootstrap method.

\section{Discussion}%
\label{sec:discussion}

\subsection{Conformity signal for low-mass central galaxies}%
\label{sub:conformity_signal_for_low_mass_central_galaxies}

\begin{figure*}
    \centering
    \includegraphics[width=0.9\linewidth]{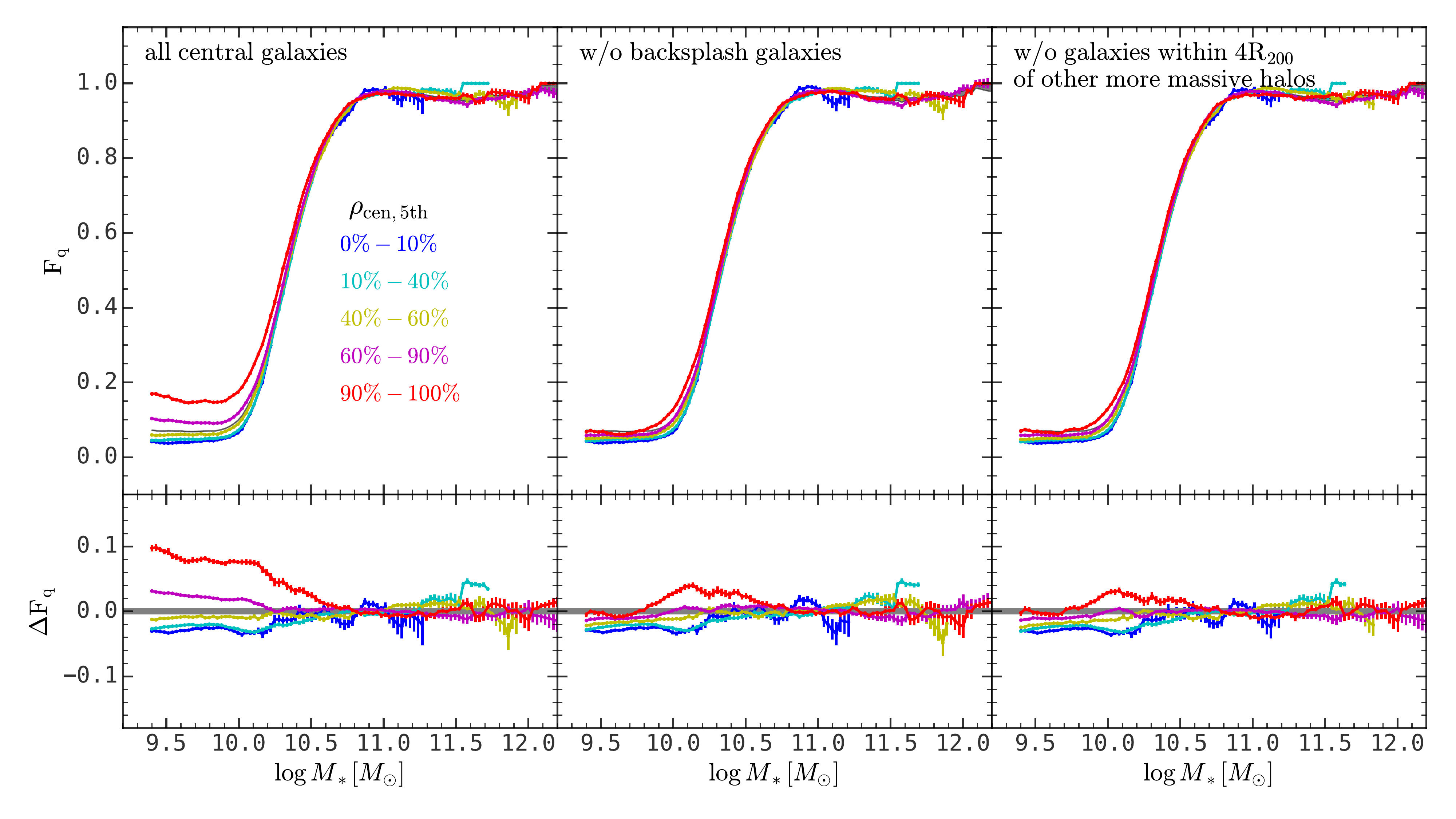}
    \caption{
        {\bf Upper panels:} The quiescent fraction as a function of stellar
        mass in different $\rho_{\rm cen, 5th}$ quantiles in the IllustrisTNG
        simulation at $z=0$. The gray filler regions are the results for all
        central galaxies. Error bars are calculated from bootstrap samples.
        {\bf Lower panels:} The residual of the quiescent fraction with respect
        to the result for all central galaxies. The first column shows the
        result for all central galaxies. The second column shows the result
        after removing all backsplash galaxies (see
        \S~\ref{sub:simulation_data}). The last column shows the result after
        removing all central galaxies within $4R_{\rm 200}$ of any other more
        massive halos. One can see that the environmental dependence of the
        quiescent fraction for low-mass central galaxies in the IllustrisTNG
        simulation can be entirely explained by backsplash galaxies. And the
        IllustrisTNG simulation cannot reproduce the signal for massive
        galaxies as we have seen in observation (see
        Fig.~\ref{fig:figure/gal_prop_cen_on_vmax_sigma_cen_1}).
    }%
    \label{fig:figure/gal_prop_cen_sigma_cen_tng300_compare}
\end{figure*}

\begin{figure*}
    \centering
    \includegraphics[width=0.9\linewidth]{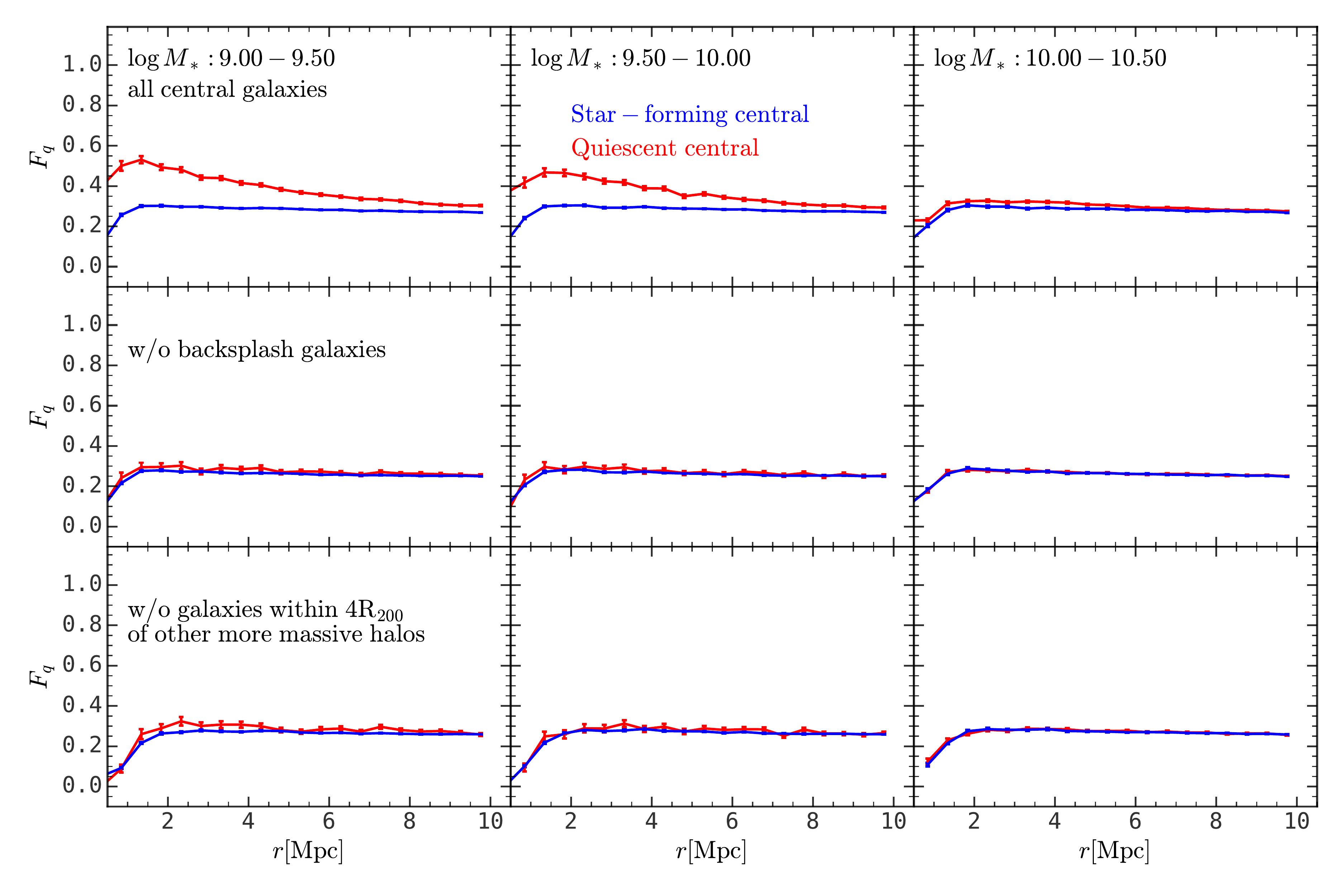}
    \caption{
        The quiescent fraction for central galaxies with $M_*\geq
        10^{9}M_{\odot}$ as a function of distance to central galaxies in three
        stellar mass bins in the IllustrisTNG simulation. Panels on the first
        row are for all central galaxies. The panels on the second row are for
        all central galaxies except backsplash ones. Panels on the third are
        for all central galaxies except those within $4R_{200}$ of any other
        more massive halos. These results also indicate that the two-halo
        conformity effect for central galaxies in the IllustrisTNG simulation
        can be entirely explained by backsplash galaxies.
    }%
    \label{fig:figure/conformity_cross_tng300}
\end{figure*}

\begin{figure*}
    \centering
    \includegraphics[width=0.9\linewidth]{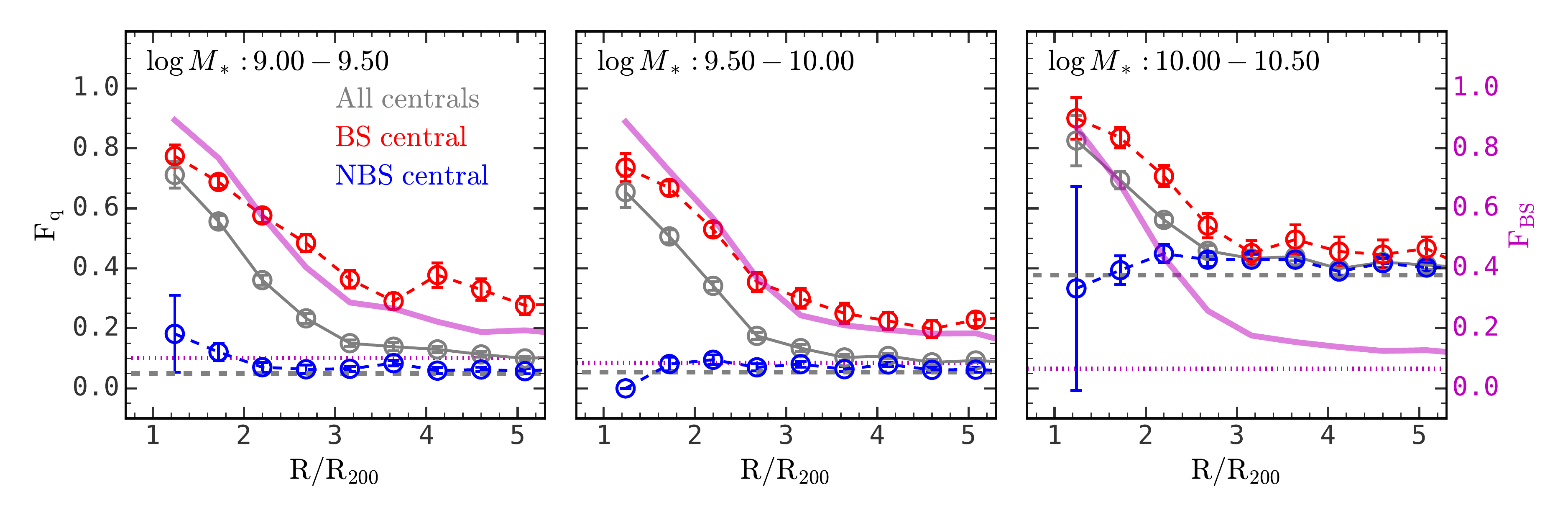}
    \caption{
        {\bf Left $y$-axis:} The quiescent fraction of central galaxies around
        dark matter halos with $M_h\geq 10^{13}M_\odot$ as a function of
        normalized distance in three stellar mass bins in the IllustrisTNG
        simulation. The gray circles show the results for all central galaxies,
        the red circles are for all backsplash galaxies, and the blue circles
        are for all non-backsplash galaxies. Error bars are estimated as the
        standard deviation of bootstrap samples. The gray horizontal dashed
        lines are the average quiescent fraction for all non-backsplash central
        galaxies in the simulation volume in three stellar mass bins. {\bf
        Right $y$-axis:} The magenta lines show the fraction of the backsplash
        galaxies among all central galaxies around halos with $M_h\geq
        10^{13}M_\odot$ in the IllustrisTNG simulation. The magenta horizontal
        dotted lines are the average fraction of backsplash galaxies among all
        central galaxies in the simulation volume in three stellar mass bins.
        This figure demonstrates that the quiescent fraction excess for central
        galaxies around massive halos in the IllustrisTNG simulation all stems
        from backsplash galaxies.
    }%
    \label{fig:figure/fq_around_cluster}
\end{figure*}

\begin{figure*}
    \centering
    \includegraphics[width=0.8\linewidth]{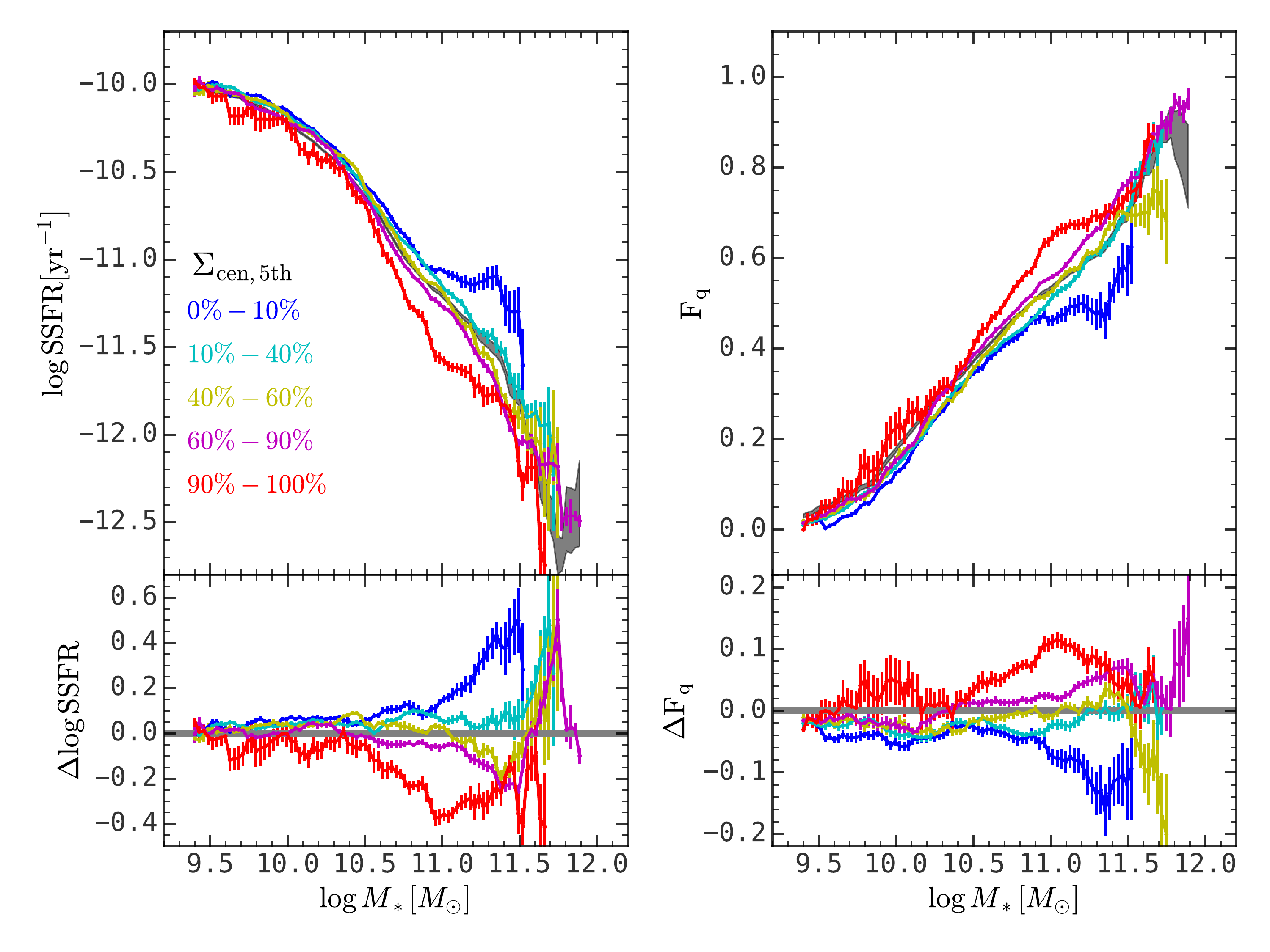}
    \caption{
        Similar to Fig.~\ref{fig:figure/gal_prop_cen_on_vmax_sigma_cen_1}. Here
        we exclude galaxies within $4R_{200}$ of any other more massive halos
        with $\Delta V < 1000\rm km/s$. One can see that the conformity signal
        for low-mass galaxies are largely eliminated, indicating that this
        effect is mainly contributed by backsplash galaxies.
    }%
    \label{fig:figure/gal_prop_cen_on_vmax_sigma_cen_1_exclud_mnbr}
\end{figure*}

Fig.~\ref{fig:figure/gal_prop_cen_on_vmax_sigma_cen_1} shows that low-mass
galaxies with $M_* \lesssim 10^{10.5}M_{\odot}$ residing in high-density
regions are more quenched than their counterparts in low-density regions. This
phenomenon is not only found in observation, but also shown in the
semi-analytical models (SAMs) and hydrodynamical simulations of galaxy
formation \citep[e.g.][]{lacernaGalacticConformityMeasured2018,
lacernaEnvironmentalInfluenceGroups2022, ayromlouPhysicalOriginGalactic2022a}.
\citet{lacernaEnvironmentalInfluenceGroups2022} uses the IllustrisTNG
simulation and the SAM of \texttt{SAG}
\citep[e.g.][]{springelPopulatingClusterGalaxies2001,
coraSemianalyticGalaxiesSynthesis2018} to demonstrate that this effect is
attributed to the vicinity of low-mass central galaxies around massive
clusters, so it disappears once these galaxies are removed. Similarly,
\citet{ayromlouPhysicalOriginGalactic2022a} find that the original SAM of
\texttt{L-GALAXIES} in \citet{henriquesGalaxyFormationPlanck2015} has no
conformity signal until a new recipe is adopted, where the ram-pressure
stripping effect can also affect the evolution of galaxies beyond the virial
radius of massive halos \citep{ayromlouGalaxyFormationLGALAXIES2021}.

Here we use the IllustrisTNG simulation to see the dependence of central-galaxy
properties on large-scale environments and explore their physical origins.
Since we intend to study similar signals in IllustrisTNG instead of performing
a quantitative comparison with observation, there is no need to include
observation effects like the redshift-space distortion effect, which can only
undermine the strength of this conformity signal. Here we define a 3D-version
large-scale environment metric of equation~\ref{eq:sigma_cen} for each central
galaxy, i.e.
\begin{equation}
    \rho_{\rm cen,5th} = \frac{3\times 5}{4\pi D_{\rm 5th}^3}
\end{equation}
where $D_{\rm 5th}$ is the distance to the 5th nearest central galaxy.

Fig.~\ref{fig:figure/gal_prop_cen_sigma_cen_tng300_compare} shows the quiescent
fraction of central galaxies as a function of stellar mass in different bins of
$\rho_{\rm cen,5th}$, where the left panel shows the result for all central
galaxies. Clearly, one can see that low-mass central galaxies with $M_*\lesssim
10^{10.5}M_{\odot}$ in dense regions are more quenched than their counterparts
in under-dense regions and the difference in the quiescent fraction is up to
$12\%$. The middle panel of
Fig.~\ref{fig:figure/gal_prop_cen_sigma_cen_tng300_compare} shows the result
after removing all backsplash central galaxies, which are central galaxies at
$z=0$ and were satellite galaxies in other halos at $z< 1$\footnote{Here we
    only consider backsplash galaxies that were satellite galaxies at $z< 1$
    since, at higher redshift, the progenitor galaxies have very low stellar
mass and suffer from resolution problems.}. Here one can see that the
conformity signal completely disappears on the whole stellar mass range.
This result suggests that the conformity signal for low-mass central
galaxies in the IllustrisTNG simulation mainly comes from the backsplash
galaxies \citep{wetzelGalaxyEvolutionGroups2014}. As suggested in
\citet{wetzelGalaxyEvolutionGroups2014}, we can also eliminate the impact
of backsplash galaxies by removing galaxies within $4R_{200}$ of any other
more massive halos, and the result is shown on the right panel of
Fig.~\ref{fig:figure/gal_prop_cen_sigma_cen_tng300_compare}. Here we can
see that the conformity signal for galaxies with $M_*\lesssim
10^{10.5}M_{\odot}$ also disappears.

Previous studies quantify the two-halo galactic conformity effect with the
quiescent fraction difference of neighboring galaxies ({\it secondary
galaxies}) around star-forming and quiescent central galaxies ({\it primary
galaxies}). \citet{lacernaEnvironmentalInfluenceGroups2022} found that galaxies
around quiescent central galaxies are more quenched than those around
star-forming central galaxies with similar stellar masses in the IllustrisTNG
simulation, and this difference is prominent up to $\sim 10$ Mpc \citep[see
also][]{ayromlouPhysicalOriginGalactic2022a}. Here we want to see if
backsplash galaxies can explain the two-halo galactic conformity signal
manifested in this way. In previous studies, the primary galaxies only include
centrals, and the secondary galaxies include both centrals and satellites. As
we argued in \S\,\ref{sec:introduction} (see also
\S\,\ref{sub:two_halo_conformity_effect_on_satellite_galaxies}), the mixing of
central and satellite galaxies makes the result hard to interpret, so here we
only include centrals for both the primary and the secondary galaxies.

Fig.~\ref{fig:figure/conformity_cross_tng300} shows the quiescent fraction of
central galaxies with $M_*\geq 10^{9}M_{\odot}$ (secondary galaxies) as a
function of distance to central galaxies in three stellar mass bins (primary
galaxies). The primary galaxies are further separated into star-forming and
quiescent populations and the results are shown in blue and red colors,
respectively. Top panels in Fig.~\ref{fig:figure/conformity_cross_tng300} show
the results for all central galaxies in the TNG simulation. Here one can see
that secondary galaxies around quiescent primary galaxies have higher quiescent
fraction by $\lesssim 20\%$, and the difference in the quiescent fraction
decreases for more massive primary galaxies and for larger separations, until
the difference disappears for the primary galaxies with $M_*\gtrsim
10^{10}M_{\odot}$ or for $r > 8$ Mpc. Panels on the second row in
Fig.~\ref{fig:figure/conformity_cross_tng300} show the result after removing
all backsplash central galaxies. Here one can see that the difference in the
quiescent fraction for star-forming and quiescent primary galaxies disappears,
indicating that the two-halo galactic conformity signal comes from these
backsplash galaxies. Finally, we tested the method suggested in
\citet{wetzelGalaxyEvolutionGroups2014} to eliminate the impact of backsplash
galaxies by removing centrals within $4R_{\rm 200}$ of any other more massive
halos, and the result is shown on the bottom panels in
Fig.~\ref{fig:figure/conformity_cross_tng300}. Again, the conformity signal
completely disappears. It is noteworthy that previous studies have shown that
low-mass central galaxies close to more massive halos are more likely to be
backsplash galaxies \citep{wetzelGalaxyEvolutionGroups2014}, which is also
consistent with the result here.

In order to further understand the distribution of backsplash galaxies around
massive halos and their contribution to the two-halo conformity signal for
central galaxies, we select all central galaxies around halos with $M_h\geq
10^{13} M_\odot$ in the IllustrisTNG simulation and calculate their quiescent
fractions and backsplash fractions in Fig.~\ref{fig:figure/fq_around_cluster}.
The quiescent fractions of central galaxies around these massive halos are
shown in gray circles as a function of normalized distance. Here one can see
prominent quiescent fraction excess for central galaxies in the vicinity of
these massive halos, which is a manifestation of the two-halo conformity effect
for central galaxies. Then, we separate these central galaxies into the backsplash
ones and non-backsplash ones and plot their quiescent fractions in red and blue
colors, respectively. Here one can see that the quiescent fraction for
non-backsplash central galaxies shows no excess anymore and coincides with the
average quiescent fraction for all non-backsplash galaxies in the simulation
volume. This indicates that the quiescent fraction excess around massive halos
in the IllustrisTNG simulation, which is responsible for the two-halo
conformity effect for central galaxies, all stems from backsplash galaxies.
Meanwhile, the right $y$-axis of Fig.~\ref{fig:figure/fq_around_cluster} shows
the fraction of backsplash galaxies among all central galaxies around these
massive halos, where the backsplash galaxies range from 20\% to 80\%. We note
that backsplash galaxies dominate (>50\%) the central galaxy population within
$2R_{\rm 200}$ around these massive systems and gradually declines to $\lesssim
20\%$ around $4R_{\rm 200}$. Eventually, they will approach the average
fraction (5\%-10\%) in the simulation volume, as shown in magenta horizontal
dotted lines. We note that those distant backsplash galaxies are not all
associated with the halos in question and, instead, from other nearby massive
halos \citep{wetzelGalaxyEvolutionGroups2014}. This figure clearly demonstrates
that the mere presence of massive halos around central galaxies cannot affect
their star formation states, and more violent within-halo environmental
processes are required. Nevertheless, it is noteworthy that non-backsplash
central galaxies in the vicinity of massive halos may suffer from some other
environmental effects, like the termination of gas accretion and the stripping
of the circum-galactic medium, as found in previous studies
\citep[e.g.][]{baheWhyDoesEnvironmental2013, behrooziMergersMassAccretion2014,
ayromlouGalaxyFormationLGALAXIES2021}.

The above results clearly demonstrate that the conformity signal in the
IllustrisTNG simulation can be entirely explained by backsplash galaxies. In
real observation, since we have no access to their halo merger trees, we can
only eliminate the impact of backsplash galaxies by removing galaxies close to
any other more massive halos. Here we remove all central galaxies within
$4R_{\rm 200}$ of any other more massive halos with $\Delta V < 1000\rm km/s$,
and Fig.~\ref{fig:figure/gal_prop_cen_on_vmax_sigma_cen_1_exclud_mnbr} shows
the resulting median SSFR and the quiescent fraction for central galaxies as a
function of stellar mass and $\Sigma_{\rm cen, 5th}$. Here one can see that the
dependence of SSFR and quiescent fraction on $\Sigma_{\rm cen,5th}$ are nearly
eliminated, indicating this conformity signal is mainly due to the presence of
backsplash galaxies. Again, we can still see a weak dependence on $\Sigma_{\rm
cen, 5th}$. This is consistent with the weak signal presented in
\citet{tinkerHaloHistoriesGalaxy2017}, in which they suggest that this signal
is attributed to the correlation between halo formation histories and the star
formation activities of central galaxies.

Finally, we want to emphasize that the results in
Fig.~\ref{fig:figure/gal_prop_cen_sigma_cen_tng300_compare} and
Fig.~\ref{fig:figure/conformity_cross_tng300} only indicate that the two-halo
conformity effect for central galaxies in IllustrisTNG can be explained by
backsplash galaxies, but cannot falsify the existence of super-halo-scale
environmental effects on low-mass central galaxies. Our results highlight the
role played by these backsplash galaxies, even though only $20\%-80\%$ of the
central galaxies around massive halos are backsplash galaxies (see
Fig.~\ref{fig:figure/fq_around_cluster}). We also emphasize the necessity to
control the contribution from these galaxies before we can study the impact of
super-halo-scale environmental effects. Finally, investigations on other models
can help us better understand the environmental effect, within-halo and
super-halo-scale, on galaxies. For example,
\citet{ayromlouPhysicalOriginGalactic2022a} performs a similar analysis on the
semi-analytical of \citet{ayromlouGalaxyFormationLGALAXIES2021}, where they
found that the two-halo conformity effect on central galaxies is still
prominent after removing backsplash galaxies from the primary sample and all
satellite galaxies are retained in the secondary sample (see their Figure 8).

\subsection{Conformity signal for massive central galaxies}%
\label{sub:conformity_signal_for_massive_central_galaxies}

\begin{figure*}
    \centering
    \includegraphics[width=0.8\linewidth]{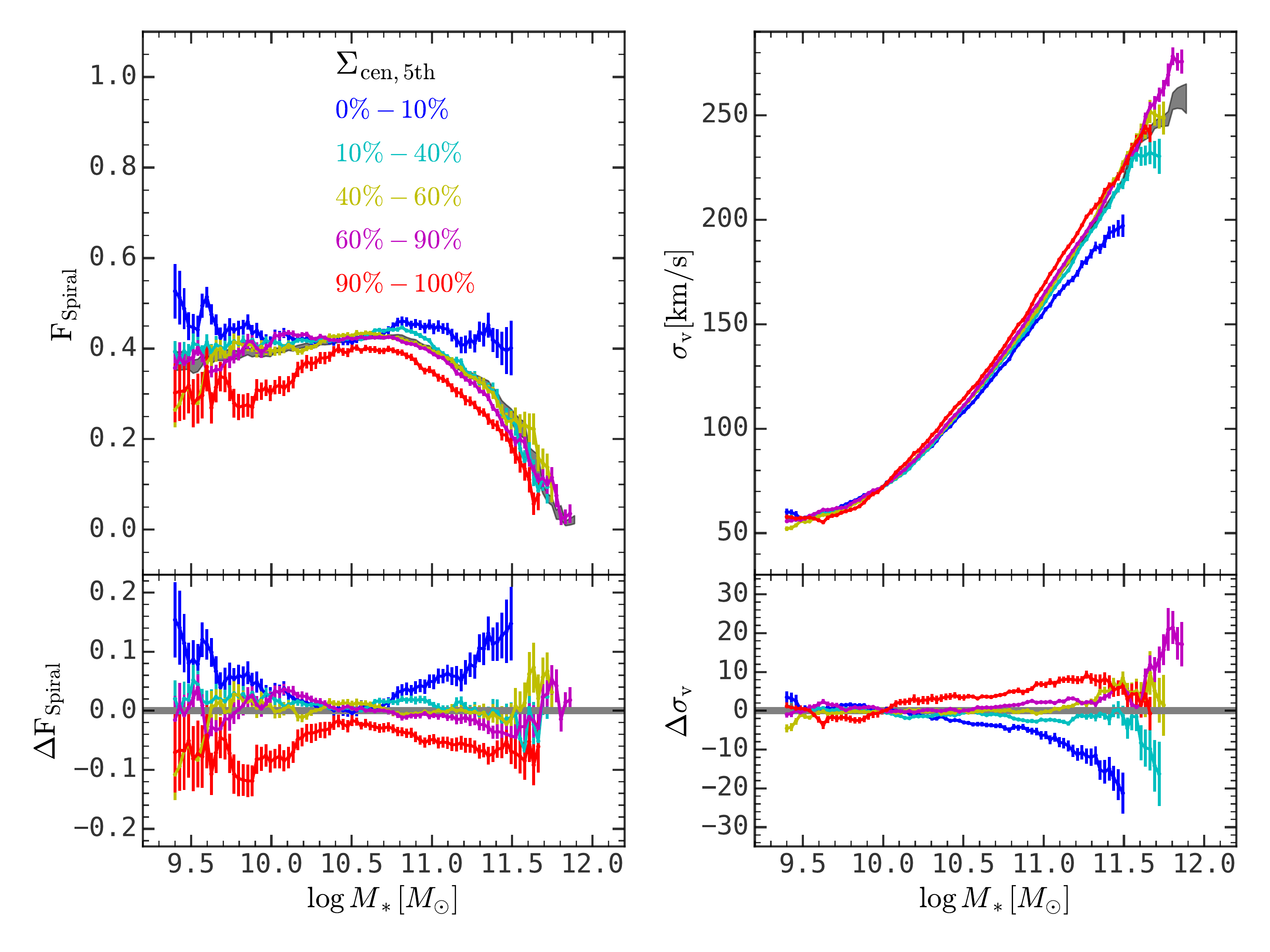}
    \caption{
        The fraction of spiral galaxies ({\bf left panel}) and the median
        central stellar velocity dispersion ({\bf right panel}) for central
        galaxies as a function of stellar mass in different $\Sigma_{\rm cen,
        5th}$ bins for SDSS galaxies. Error bars are calculated from bootstrap
        samples. And the lower panels show the residual with respect to the
        results for all of the central galaxies at given stellar mass.
    }%
    \label{fig:figure/gal_prop_cen_on_vmax_sigma_cen_2}
\end{figure*}

From Fig.~\ref{fig:figure/gal_prop_cen_on_vmax_sigma_cen_1}, we can also see
that massive central galaxies with $M_*\gtrsim 10^{11}M_{\odot}$ show strong
dependence on $\Sigma_{\rm cen, 5th}$, where those in low-density regions are
more star-forming than their counterparts in dense regions. Moreover, this
dependence is still there after the elimination of the backsplash galaxies (See
Fig.\ref{fig:figure/gal_prop_cen_on_vmax_sigma_cen_1_exclud_mnbr}), indicating
that this effect cannot be attributed to these backsplash galaxies. A related
trend was also pointed out in \citet{tinkerHaloHistoriesGalaxy2018}, where they
focused on star-forming galaxies and found that star-forming central galaxies
preferentially live in low-density regions.

Recently, \citet{zhangMassiveStarformingGalaxies2022} found that star-forming
massive central galaxies prefer to reside in low-mass halos than their
quiescent counterparts based on weak lensing measurements \citep[see
also][]{mandelbaumDensityProfilesGalaxy2006, moreSatelliteKinematicsIII2011,
    rodriguez-gomezMergerRateGalaxies2015, mandelbaumStrongBimodalityHost2016,
zhangHostsTriggersAGNs2021}. Consequently, these star-forming central galaxies
are less clustered than those quiescent ones (see also Figure 6 of their
paper), as expected since massive halos have higher clustering strength
\citep{moAnalyticModelSpatial1996}. This result qualitatively agrees with
findings in this paper.

To understand the origin of this signal, we investigated their morphology and
kinematics, and the results are presented in
Fig.~\ref{fig:figure/gal_prop_cen_on_vmax_sigma_cen_2}. On the left panel, we
can see that these massive galaxies ($M_*\gtrsim 10^{11}M_{\odot}$) in
low-density regions are preferentially spiral compared with massive galaxies in
dense regions. Meanwhile, from the right panel, one can see that the median
central stellar velocity dispersion is also lower for massive galaxies in
sparse regions than their counterparts in dense regions by $\sim 15\%$,
indicating relatively low-mass bulges and central black holes. All these
results together suggest a simple explanation for the conformity signal we
observed here. Galaxies in sparse regions experienced fewer merger events
\citep{fakhouriEnvironmentalDependenceDark2009,
    jianENVIRONMENTALDEPENDENCEGALAXY2012,
kampczykEnvironmentalEffectsInteraction2013}, thus their spiral morphology
keeps intact and the growths of their bulge and central black hole are
suppressed. Consequently, they are less likely to be quenched. This explanation
is also supported by \citet{postiPeakStarFormation2019}, where they found that
spiral galaxies can convert baryons to stars much more efficiently than
galaxies with non-spiral morphology \citep[see
also][]{postiDynamicalEvidenceMorphologydependent2021,
mancerapinaImpactGasDisc2022}.

It is noteworthy that the IllustrisTNG simulation cannot reproduce the
conformity signal for massive galaxies. In the IllustrisTNG model, the star
formation activities of massive central galaxies are mostly determined by the
integrated kinetic energy released by the AGN feedback process, or
equivalently, the black hole mass \citep{terrazasRelationshipBlackHole2020,
    xuCriticalStellarCentral2021, piotrowskaQuenchingStarFormation2022,
bluckQuenchingGalaxiesBulges2022a}. Meanwhile, the mass of central massive
black holes strongly correlates with the total stellar mass of their host
galaxies \citep{terrazasRelationshipBlackHole2020,
habouzitSupermassiveBlackHoles2021}. Consequently, the star formation
activities of massive central galaxies ($\gtrsim 10^{11} M_{\odot}$) are mostly
quenched due to the possession of massive black holes and have no dependence on
their large-scale structure. In observation, the star formation activities of
central galaxies also mostly depend on the mass of their central black hole
\citep{piotrowskaQuenchingStarFormation2022}. However, observational results
reveal a tight relation between black hole masses and bulge masses
\citep[e.g.][]{magorrianDemographyMassiveDark1998,
kormendySupermassiveBlackHoles2011, kormendyCoevolutionNotSupermassive2013},
and the relation between bulge masses and the stellar masses of host galaxies
is not tight due to a broad distribution of the bulge-to-total mass ratio (B/T)
\citep[e.g.][]{davisBlackHoleMass2018}, where massive star-forming central
galaxies tend to have lower B/T than their quiescent counterparts
\citep{zhangMassEnvironmentDrivers2021a, shiColdGasMassive2022}.

\subsection{Two-halo conformity effect on satellite galaxies}%
\label{sub:two_halo_conformity_effect_on_satellite_galaxies}

In this work, we only consider central galaxies in the secondary galaxy sample,
while previous studies on the two-halo conformity effect include both central
and satellite galaxies. We emphasize that it is helpful to separate these two
populations of galaxies for a better understanding of the underlying physical
drivers of the two-halo conformity effect.

Before one can study the impact of the two-halo conformity effect on satellite
galaxies\footnote{Satellite galaxies can be within or beyond halo virial radius
due to the anisotropic spatial distribution of halos.}, it is necessary to
carefully control the within-halo environmental effect from their own host
halos. Previous studies, from both observations and simulations, reveal that
the within-halo environmental effect severely affects the properties of
satellite galaxies by shutting down the new gas replenishment, stripping the
existing cold gas content, and quenching their star formation activities. And
the major determinants for the within-halo environmental effect are halo mass
and halo-centric distance \citep[e.g.][]{wooDependenceGalaxyQuenching2013,
wangELUCIDIVGalaxy2018}. Consequently, in order to study the impact of the
two-halo conformity on satellite galaxies, one must control the influence of
their own host halos, which is beyond the scope of this paper.

Moreover, previous studies found that quiescent central galaxies prefer to live
in more massive halos than their star-forming counterparts
\citep[e.g.][]{mandelbaumStrongBimodalityHost2016,
zhangMassiveStarformingGalaxies2022}, which is related to the one-halo
conformity effect where halos with quiescent central galaxies prefer to host
quiescent satellites \citep{weinmannPropertiesGalaxyGroups2006}. Combined with
the fact that more massive halos are more clustered
\citep[e.g.][]{moAnalyticModelSpatial1996}, satellite galaxies can exhibit the
two-halo conformity signal without any new physical processes introduced.

\section{Conclusion}%
\label{sec:conclusion}

Previous studies show that the star formation activities of central galaxies
are primarily determined by their stellar mass. In addition, their star
formation activities are also found to be in correlation over several Mpcs.
This is known as the two-halo galactic conformity effect, and its physical
origin is still under debate. In this paper, we studied one manifestation of
the two-halo galactic conformity effect by investigating the dependence of the
star formation activities of central galaxies on the large-scale environment.
Our main results are summarized as follows:
\begin{enumerate}
    \item We proposed a novel large-scale environment metric, $\Sigma_{\rm cen,
        5th}$, which is inferred from the distance to the fifth nearest central
        galaxy. This metric measures the environment from $\sim 1$ Mpc to
        $\gtrsim 10$ Mpc, with a median value of $\sim 4$ Mpc. With this
        environment metric, we found two kinds of minor but significant
        dependence on $\Sigma_{\rm cen, 5th}$. The first one is that central
        galaxies in the highest-density regions are more quenched by $\sim 10$
        percent than those in the lowest-density regions, and the median SSFR
        is lower by $\sim 0.2$ dex. The second one is that massive central
        galaxies in the lowest-density regions are less quenched by $\gtrsim
        15$ percent than those in the highest-density regions, and the median
        SSFR is also higher by $\sim 0.4$ dex (see
        Fig.~\ref{fig:figure/gal_prop_cen_on_vmax_sigma_cen_1}).

    \item We found a similar trend in the IllustrisTNG simulation where
        low-mass central galaxies are more quenched in dense regions than those
        in sparse regions by $\lesssim 12$ percent. We found that this signal
        is mainly contributed by backsplash galaxies, which are central
        galaxies that once were satellites of other halos, so the environmental
        dependence of the quiescent fraction for low-mass central galaxies
        disappears once these backsplash galaxies are removed. The impact of
        these backsplash galaxies can also be eliminated by removing central
        galaxies in the vicinity of more massive halos in the IllustrisTNG
        simulation. By applying the same procedure to the observational data,
        we found that the environmental dependence of the quiescent fraction
        for low-mass central galaxies is largely reduced, suggesting the
        conformity signal for low-mass central galaxies in observation also
        comes from backsplash galaxies (see
        Fig.~\ref{fig:figure/gal_prop_cen_sigma_cen_tng300_compare} and
        Fig.~\ref{fig:figure/gal_prop_cen_on_vmax_sigma_cen_1_exclud_mnbr}).

    \item We found that the two-halo galactic conformity effect for central
        galaxies manifested by the quiescent fraction difference of neighboring
        central galaxies (secondary galaxies) around star-forming and quiescent
        central galaxies (primary galaxies) decreases with increasing the
        stellar mass of primary galaxies and increasing the separation between
        the primary and the secondary galaxies in the IllustrisTNG simulation,
        which is consistent with previous results. Most importantly, we found
        that this signal can be entirely explained by backsplash galaxies,
        where this signal completely disappears once backsplash galaxies are
        removed (see Fig.~\ref{fig:figure/conformity_cross_tng300}).

    \item Massive central galaxies in the lowest-density regions are not only
        more star-forming, but also have a higher fraction of spiral morphology
        by $\sim 15$ percent and lower median central stellar velocity
        dispersion by $\sim 10$ percent. These results suggest a simple
        scenario that these galaxies in sparse regions on average have
        experienced fewer major merger events, and as a consequence, they have
        higher disk frequency, lower central stellar velocity dispersion, less
        massive central black holes. Hence they are more likely to survive from
        star formation suppression processes, including merging-quenching,
        morphological quenching and AGN feedback that are closely related to a
        massive classical bulge and a massive black hole (see
        Fig.~\ref{fig:figure/gal_prop_cen_on_vmax_sigma_cen_2}).

\end{enumerate}

\section*{Acknowledgements}

The authors thank the anonymous referee for their helpful comments that
improved the quality of the manuscript. This work is supported by the National
Science Foundation of China (NSFC) Grant No. 12125301, 12192220, 12192222, and
the science research grants from the China Manned Space Project with NO.
CMS-CSST-2021-A07.

The authors acknowledge the Tsinghua Astrophysics High-Performance Computing
platform at Tsinghua University for providing computational and data storage
resources that have contributed to the research results reported within this
paper.

Funding for the SDSS and SDSS-II has been provided by the Alfred P. Sloan
Foundation, the Participating Institutions, the National Science Foundation,
the U.S. Department of Energy, the National Aeronautics and Space
Administration, the Japanese Monbukagakusho, the Max Planck Society, and the
Higher Education Funding Council for England. The SDSS Web Site is
http://www.sdss.org/.

The SDSS is managed by the Astrophysical Research Consortium for the
Participating Institutions. The Participating Institutions are the American
Museum of Natural History, Astrophysical Institute Potsdam, University of
Basel, University of Cambridge, Case Western Reserve University, University of
Chicago, Drexel University, Fermilab, the Institute for Advanced Study, the
Japan Participation Group, Johns Hopkins University, the Joint Institute for
Nuclear Astrophysics, the Kavli Institute for Particle Astrophysics and
Cosmology, the Korean Scientist Group, the Chinese Academy of Sciences
(LAMOST), Los Alamos National Laboratory, the Max-Planck-Institute for
Astronomy (MPIA), the Max-Planck-Institute for Astrophysics (MPA), New Mexico
State University, Ohio State University, University of Pittsburgh, University
of Portsmouth, Princeton University, the United States Naval Observatory, and
the University of Washington.

\section*{Data availability}

The data underlying this article will be shared on reasonable request to the
corresponding author. The computation in this work is supported by the HPC
toolkit \specialname[hipp] at \url{https://github.com/ChenYangyao/hipp}.

\appendix

\bibliographystyle{mnras}
\bibliography{bibtex.bib}

\bsp	
\label{lastpage}
\end{document}